| | |
|---|---|
| **Abstract** | Halophilic microorganisms have long been known to survive within the brine inclusions of salt crystals, as evidenced by the change in color for salt crystals containing pigmented halophiles. However, the molecular mechanisms allowing this survival has remained an open question for decades. While protocols for the surface sterilization of halite (NaCl) have enabled isolation of cells and DNA from within halite brine inclusions, "-omics" based approaches have faced two main technical challenges: (1) removal of all contaminating organic biomolecules (including proteins) from halite surfaces, and (2) performing selective biomolecule extractions directly from cells contained within halite brine inclusions with sufficient speed to avoid modifications in gene expression during extraction. In this study, we tested different methods to resolve these two technical challenges. Following this method development, we then applied the optimized methods to perform the first examination of the early acclimation of a model haloarchaeon (Halobacterium salinarum NRC-1) to halite brine inclusions. Examinations of the proteome of Halobacterium cells two months post-evaporation revealed a high degree of similarity with stationary phase liquid cultures, but with a sharp down-regulation of ribosomal proteins. While proteins for central metabolism were part of the shared proteome between liquid cultures and halite brine inclusions, proteins involved in cell mobility (archaellum, gas vesicles) were either absent or less abundant in halite samples. Proteins unique to cells within brine inclusions included transporters, suggesting modified interactions between cells and the surrounding brine inclusion microenvironment. The methods and hypotheses presented here enable future studies of the survival of halophiles in both culture model and natural halite systems. |
| **Keywords** | Halobacterium, halophile, LC-MS, proteomics, halite (NaCl) |

# 1. Introduction

Extremely halophilic archaea are adapted to the highest possible salinity conditions, thriving in saturated brines. Microbial diversity in such environments is highly reduced and tends to be dominated by haloarchaea (Oren, 2006). Saturated brines are characterized by a low dissolved oxygen content at the surface despite their contact with air (Sherwood et al., 1991). The surface of saturated brines is often exposed to high temperatures and high levels of solar radiation (Oren, 2006; Merino et al., 2019), which can lead to evaporation. During evaporation events, the total contents of the brine, including salts, organics, some atmospheric gases, and any microorganisms present, are trapped within inclusions in the salt crystals. Viable halophilic bacteria and archaea have been isolated from halite (NaCl) of various ages, from months to years (Norton and Grant, 1988; Gramain et al., 2011; Huby et al., 2020), to geologically relevant timescales with the age of the primary halite given by the surrounding geological matrix. Some studies have even reported the observation or isolation of microorganisms from halite dated to several hundreds of millions of years (Vreeland et al., 2000; Schreder-Gomes et al., 2022). These observations have raised two important questions: (1) are these truly "ancient" cells? (2) what physiological changes are required for microbial cells to remain viable within halite brine inclusions?

The first question has been studied in greater detail, and while many questions still remain, these studies provided valuable insights into the potential for surface-contaminant microorganisms (Graur and Pupko, 2001; Maughan et al., 2002; Nickle et al., 2002). In response to these criticisms, appropriate surface sterilization and cleaning procedures for microbial cultivation and DNA isolation for micro-biodiversity studies from dissolved ancient halite (Gramain et al., 2011; Sankaranarayanan et al., 2011) were developed. These approaches enabled more accurate identification of only those microorganisms trapped within halite inclusions. Another opened question is whether the age determined from the geological context in which primary halite crystals were found can accurately be applied directly to the microorganisms in the brine inclusions (Hazen and Roedder, 2001). Dissolution and recrystallization of halite over geological time scales may contribute to the presence of more modern microorganisms (see discussion in Winters et al., 2015 and references therein). No known microbial survival mechanism to date can account for cell viability over millions of years. Some studies have examined the possibility that microbial cell-like biomorphs are preserved within brine inclusions (Nims et al., 2021).

An alternative approach is to experimentally confirm how microbial life is supported within brine inclusions over durations known to support viable microorganisms. Determining the cellular functions expressed by viable halophiles within halite brine inclusions requires the direct isolation of biomolecules such as proteins and nucleic acids for "-omics" analyses, including proteomics and transcriptomics. However, significant technical challenges are presented by the closed system of brine inclusions within halite. Accessing the contents of brine inclusions by rapid crystal dissolution leads to cell lysis by osmotic shock and biomolecule degradation, while the time required for gradual crystal dissolution preserving cellular integrity is also sufficient for alterations in the transcriptomes and proteomes of viable cells away from their state within the brine inclusions. In addition, the direct extraction of biomolecules from brine inclusions involves a large amount of NaCl, which is incompatible with direct mass spectrometry analysis due to the lack of sufficient desalination steps in standard protocols. Although current hypotheses concerning the modifications of halophile physiology during entrapment within halite brine inclusions include a change to anaerobic metabolism (Winters et al., 2015) along with potential for cell envelope modifications depending on conditions (Fendrihan et al., 2012; Kixmüller and Greie, 2012), these remain largely unverified at the molecular level due to these experimental challenges. New methods are needed to isolate proteins and other biomolecules directly from halite brine inclusions, without allowing the microorganisms to alter their gene expression during salt dissolution and processing. The development of a new analytical workflow compatible with "-omics" analyses is best conducted with a known model organism for which there is a large repertoire of physiological studies and multi-omics data for liquid cultures exposed to a range of conditions applicable to halite brine inclusions (different oxygen availabilities, salinities, nutrient availabilities, etc.). For these reasons, we choose the model haloarchaeon *Halobacterium salinarum*, the type strain of the Halobacteriales family (Gruber et al., 2004; Oren et al., 2009).

*Halobacterium salinarum* is an appropriate model as it is found both in contemporary NaCl-saturated aqueous environments such as Great Salt Lake (Post, 1977) and has been detected in halite and closely related to isolates from ancient salt deposit (Mormile et al., 2003) The red pigmentation of *H. salinarum* strain is directly correlated with transmembrane proteins (bacteriorhodopsin and halorhodopsin) and carotenoids (bacterioruberin) also used by cells as antioxidants (Eichler, 2019). The accumulation of high intracellular concentrations of potassium ($K^+$) and chloride ($Cl^-$) (Engel and Catchpole, 2005), needed to maintain osmotic homeostasis, induces both biochemical adaptations (in the form of a strongly acidic proteome) as well as technical challenges to desalt cellular extracts. *Halobacterium salinarum* survives in changing environments by using a complex variable energetic metabolism with a preference for aerobic respiration but is capable of switching to phototrophy via bacteriorhodopsin, arginine fermentation or anaerobic respiration using dimethyl sulfoxide (DMSO) and trimethylamine oxide (TMAO) if available (Hartmann et al., 1980; Müller and DasSarma, 2005; Falb et al., 2008; Gonzalez et al., 2009). This type of switch occurs under low oxygen conditions such as during increasing salinity linked to water evaporation. Moreover, according to Orellana et al. (2013), *H. salinarum* also seems to be able to use glycerol derived from microalgae of the genus *Dunaliella* as a source of carbon under certain specific conditions (in liquid co-culture condition with viable *Dunaliella* with high illumination over a set diurnal cycle and nitrate-limiting conditions). Established multi-omics protocols for liquid cultures of *H. salinarum* have permitted analyses of the cellular responses to a broad range of environmental conditions, including variations in NaCl (Leuko et al., 2009), pH (Moran-Reyna and Coker, 2014), oxygen (Schmid et al., 2007), and temperature (Coker et al., 2007), all relevant to halite brine inclusions. This existing knowledge base far exceeds that developed to date for halophiles isolated directly from ancient halite.

Here we present an efficient new analytical method for the study of microorganisms within halite, validated using *H. salinarum* entrapped in laboratory-grown halite to probe the question of what physiological changes are required for microbial cells to remain viable within halite brine inclusions, focusing on the initial phase of halite entrapment. The developed workflow includes removal of not only surface-attached cells and nucleic acids but also proteins, with subsequent extraction and desalting of proteins directly from brine inclusions compatible with mass spectrometry analyses. Applying these methods, we determined the acclimation of *H. salinarum* cells to inclusions at molecular level within laboratory-grown halite by analyzing the differences in the expressed proteome prior to evaporation and 2 months after culture evaporation. Analyses focused on the characterization of cellular activity compared to stationary cells not trapped within halite, as well as the interactions of cells with halite brine inclusion environment.

## 2. Materials and methods

All reagents used were analytical grade and suitable for molecular biology.

### 2.1. Strain and culture conditions

*Halobacterium salinarum* strain NRC-1 (JCM 11081) was grown under oxic conditions in autoclaved complex medium (CM: 4.28 M NaCl, 81 mM $MgSO_4 \cdot 7H_2O$, 27 mM KCl, 10 mM trisodium-citrate·$2H_2O$, 1% (w/v) peptone Oxoïd® LP0034, pH adjusted to 7.4) following Oesterhelt and Stoeckenius (1974) at 37°C, 180 rpm in glassware washed and then rinsed multiple times in MilliQ® water with vigorous shaking to remove all traces of detergents that can inhibit the growth of haloarchaea. Growth was monitored by spectrophotometry at 600 nm ($OD_{600}$). Cultures for crystallizations and protein extractions were grown to stationary phase ($OD_{600}$ = 1.0–1.6) avoiding decline phase to approximate the physiological condition of haloarchaea under natural conditions during evaporation.

## 2.2. Laboratory-grown halite

### 2.2.1. Internally inoculated halite

Modeling entombment of haloarchaea within halite was already done in laboratory by Norton and Grant (1988), Gramain et al. (2011) and Kixmüller and Greie (2012). In this study, laboratory-grown halite crystals were produced following a modified version of protocols of Fendrihan et al. (2012), by adding nutrients to their Tris-buffered NaCl solution (TN buffer; 100 mM Tris–HCl pH 7.4, 4.28 M NaCl) to simulate organic matter and nutrients in the natural environment just prior to halite precipitation (Winters et al., 2015). Briefly, *H. salinarum* cells in stationary growth phase were harvested by centrifugation at 7,500g, 10 min, 20°C and the growth medium removed by washing with sterile TN buffer. Cells were then resuspended in sterile TNPA buffer (TN buffer with 1% (w/v) Oxoïd® peptone LP0034 and 0.5% (w/v) L-Arg HCl, adjusted pH 7.4) with a ratio of 10 mL TNPA buffer per 500 mg of cells (wet weight, equal to $9.6 \times 10^{11}$ cells). Then 20 mL TNPA crystallization buffer containing $3.2 \times 10^{11}$ cells was evaporated in each sterile 90/14.3 mm Petri dish in a 37°C with a 12 h:12 h light:dark photoperiod (66.75 µmol photons.$m^{-2}$.$s^{-2}$, verified by a Li-250A, Li-Cor Inc., Germany) to model natural brine-surfaces. Complete evaporation and drying of precipitated halite were obtained after ~22 days, followed by a further 60 days (2 months) of incubation to study the early phase of *H. salinarum* entombment within halite brine inclusions.

### 2.2.2. Externally inoculated halite

To produce halite with *H. salinarum* cells localized exclusively at the halite surface, 20 mL of sterile TN buffer was first evaporated as described above until complete drying (~22 days). The resultant crystals were then collected aseptically using sterilized forceps and the surfaces of each crystal inoculated with $9 \times 10^9$ *H. salinarum* cells in stationary phase, applied as a highly concentrated cell solution after centrifugation (7,500g, 10 min, room temperature, RT°C). Crystals were inoculated on the largest faces (designated here as "top" and "bottom") by first inoculating the "top" face with $4.5 \times 10^9$ cells drop-by-drop and spreading out using a sterile inoculating loop. Halite were then dried overnight at 37°C, followed by the "bottom" face the next day with the same protocol.

## 2.3. Scanning electron microscopy

Observations and analyses of halite crystals were performed by first attaching the crystals directly to aluminum supports with carbon tape followed by carbon thin coating. Observations were performed using a Zeiss Ultra 55 field emission gun scanning electron microscope (SEM-FEG) equipped with a Brucker Energy dispersive X-ray spectroscopy (EDX) QUANTAX detector. Secondary and backscattered electron images and EDX analyses and maps were obtained at 10 kV and a working distance of 7.5 mm.

## 2.4. Post-evaporation cellular viability tests

To assess cell viability after halite inclusion, salt crystals were weighted to determine the appropriated NaCl concentration required for the complex medium to obtain a final concentration of 4.28 M after halite dissolution as previously described by Gramain et al. (2011). Halite and complex medium were then incubated at 37°C, 180 rpm. After crystal dissolution, culture $OD_{600}$ was measured by spectrophotometry. Pigmented (red) cultures reaching a normal stationary phase culture ($OD_{600}$ > 1.0) were classified as viable and cultures showing no increase in $OD_{600}$ after 1 month ($OD_{600}$ < 0.1) were classified as non-viable.

## 2.5. Halite surface cleaning (removal of cells and proteins)

### 2.5.1. Cold atmospheric plasma treatment

Cold atmospheric plasmas (i.e., weakly ionized gases) were generated were generated using either a dielectric barrier device (DBD) or an atmospheric pressure plasma jet (APPJ). They were polarized to the high voltage using two types of electrical power supply: (1) an alternative current (AC) generator composed of a function generator (ELC, GF467AF) and a power amplifier (Crest Audio, CC5500) as well as (2) a pulse generator (RLC electronic Company, NanoGen1 model) coupled with a high voltage power supply (Spellman company, SLM 10 kV 1,200 W model). DBD and APPJ were supplied with different a carrier gas (helium or argon) with/without oxygen. Hence several plasma conditions were performed to produce active species (radicals, reactive oxygen species but also electrons and photons). The **Table 1** details the experimental plasma conditions that were investigated as well as the resulting proteolysis to assess the efficiency of protein removal.

| Plasma source | Elec.parameters | | | Gas & Flow rate | Proteo-lysis |
|---|---|---|---|---|---|
| | A | F | $D_C$ | | |
| DBD (AC) | 7 kV | 500 Hz | - | He (1 slm) | 0.00 % |
| | | | | He-$O_2$ (1 slm, 150 sccm) | 0.00 % |
| | | | | He-$O_2$ (1 slm, 3 sccm) | 0.00 % |
| | | | | Ar (1 slm) | 0.00 % |
| APPJ (AC) | 7 kV | 700 Hz | - | He (6 slm) | 0.00 % |
| APPJ (Pulse) | 7 kV | 700 Hz | 1% | He (6 slm) | 29.30% |
| | | | | He (6 slm, 100 sccm) | 39.10% |

*Table 1. Experimental parameters of cold plasma sources used to remove surface-bound proteins.*

### 2.5.2. Chemical treatments

Previous studies applied surface cleaning protocols developed for removal of cells and nucleic acids, as detailed in **Supplementary Table 1-1**. To adapt these protocols for the removal of surface-bound proteins, we modified the protocol of Sankaranarayanan et al. (2011) using shorter time for baths to avoid dissolution of smaller laboratory-grown crystals. Briefly, crystals were incubated in successive 5 min baths: 4.00%–4.99% NaOCl (Honeywell Research Chemicals; avoiding commercial bleach solutions that induced rapid and near-complete crystal dissolution) in saturated NaCl and 10 M NaOH in saturated NaCl supplied or not with 10 N HCl followed by 100 mM $Na_2CO_3$ in saturated NaCl. A sterile saturated NaCl was used to wash off each treatment solution between each successive bath, testing both passive wash baths and active-spray washes using a hemolytic pump.

## 2.6. Protein extraction and preliminary desalting

We optimized a protocol for protein extraction directly from halite crystals using TRIzol Reagent™ (Ambion, Life Technologies). TRIzol allows for the sequential separation of RNA and DNA prior to protein extraction (see **Supplementary Figure 2-1**). The protocol described below was based on both the manufacturer's protocol and that of Kirkland et al. (2006) used for protein extractions from liquid cultures of Haloferax volcanii. All steps described were performed with autoclaved glassware to avoid any organics contamination from plastics (bench top protocol see **Supplementary Information Section 2**). All solvents used for protein extractions were HPLC-grade and suitable for mass spectrometry analysis. Crystals were fully immersed in 5 mL of TRIzol reagent in 30 mL glass centrifuge tube and crushed using autoclaved glass stir-rod. After 20 min incubation at 60°C, total RNA was extracted from the resulting cell lysate by adding 1 mL of 100% chloroform, incubating for 5 min at room temperature followed by phase separation by centrifugation (10,000g, 20 min, 4°C). The chloroform-containing aqueous phase was removed with autoclaved glass Pasteur pipet and 1.5 mL of 100% ethanol was added to precipitate DNA. After 3 min at room temperature, supernatant was collected by centrifugation (2,000g, 10 min, 4°C). Proteins were precipitated with 7.5 mL of 100% isopropanol and collected by centrifugation (10,000g, 20 min, 4°C). The resulting protein pellet was washed twice to remove phenol traces using 5 mL of 0.3 M guanidine-HCl in 95% ethanol to denature the proteins, followed by a third wash step with 5 mL of 100% ethanol to remove any residual guanidine-HCl. Each wash step was performed by incubating the protein pellet in the solution for 20 min at room temperature followed by centrifugation (10,000g, 20 min, 4°C). Protein desalting was accomplished by two successive protein precipitations with 2 mL of 100% glacial acetone (−20°C) and centrifugation (10,000g, 20 min, 4°C). After acetone removal, the pellet was completely dried under laminar flow. Proteins were then solubilized in 1 M $NaHCO_3$ with 0.1% SDS at room temperature for 2 days and quantified with a bicinchoninic acid (BCA) proteins assay (Pierce) using either bovine serum albumin (BSA) standards concentration from manufacturer's instruction for quantitative mass spectrometry or adapted BSA standard concentrations for low protein concentrations (see **Supplementary Information Section 3**) for evaluating removal of halite surface-bound proteins. Proteolysis rate was determined by proteins quantity comparison with and without treatments.

Total proteins were extracted from *H. salinarum* cultures in stationary growth stage using a similar procedure. For this, $2.0 \times 10^{10}$ cells from liquid cultures were pelleted by centrifugation (7,500g, 10 min, 20°C) and the cell pellets directly resuspended in 5 mL TRIzol prior to following all steps described above for halite samples. After solubilization of the protein pellet, solubilization with 1 M $NaHCO_3$ with 0.1% SDS at room temperature, protein quantification was performed using the BCA assay (Pierce) protein assays as per the manufacturer's instructions.

## 2.7. iTRAQ® isobaric labeling and mass spectrometry

Aliquots of 100 μg of proteins for each sample condition and replicate were reduced using 2 mM of tris-(2-carboxyethyl) phosphine (TCEP) at 37°C for 1 h, and alkylated with 5 mM of iodoacetamide 30 min in the dark at RT°C prior to digestion with 5 μg of trypsin Gold (Promega) for 15 h at 37°C. After digestion, additional desalting of peptides was done directly by solid phase extraction (SPE) using C18 cartridges (Sep-Pak C18 Plus Short 400 mg Sorbent, Waters). The resulting peptides were dried by speed-vac and resuspended in tetraethylammonium bromide (TEAB) 0.5 M prior to labeling. iTRAQ® labeling was performed according manufacturer's instructions (Applied Biosystems). Briefly, each of the iTRAQ® isobaric labeling reagents were reconstituted with isopropanol and then added to the 50 μg of protein digest (113, 114, 115, and 116 iTRAQ® isobaric labels for proteins from liquid controls and 117, 118, 119, and 121 for halite brine inclusions protein extractions). After 2 h at room temperature, samples were desalted again with C18 SPE. The labeled peptides eluted were then dried by speed-vac and resuspended in 2% acetonitrile, 98% $H_2O$ with 0.1% formic acid (see **Supplementary Information Section 4** for additional details).

Labeled peptide samples were analyzed by mass spectrometry as previously described (Pinel-Cabello et al., 2021) on a Q Exactive HF tandem mass spectrometer (Thermo Scientific) coupled to an UltiMate 3000 Nano LC System. Peptides were desalted online on an AcclaimPepmap100 C18 precolumn (5 μm, 100 Å, 300 μm i.d. × 5 mm) and further resolved on a nanoscale AcclaimPepmap100 C18 column (3 μm, 100 Å, 75 μm i.d. × 500 mm) at a flow rate of 200 nl.min$^{−1}$ using a 120-min gradient of 4%–32% acetonitrile. A Top 20 strategy was applied in data dependent acquisition mode. Full scan mass spectra were acquired from 350 to 1800 m/z at a resolution of 60,000 with an automatic gain control (AGC) target set at $3 \times 10^6$ ions. MS/MS fragmentation was initiated when the ACG target reached 105 ions with an intensity threshold of $9 \times 10^4$. Only precursor ions with potential charge states of 2$^+$ and 3$^+$ were selected for fragmentation applying a dynamic exclusion time of 10 s.

## 2.8. Mass spectrometry data analyses

### 2.8.1. Protein identification

Protein identifications were performed using PEAKS® X-Pro software (64 bits version, 2020, BioInformatics solutions). It allows database search assisted de novo sequencing against the protein coding sequences from *H. salinarum* NRC-1 (8,533 entries from NCBI, download date 2021/08/03). Spectral peptides matching was carried out with the following parameters: (1) mass tolerance of 10 ppm on the parent ion, (2) mass tolerance of 0.005 Da for fragment ions from MS/MS, (3) carbamidomethylated Cys (+57.0215) and iTRAQ® isobaric tag Lys and N-terminal (+304.2054) as fixed modifications; and (4) oxidized Met (+15.9949), deamidated Asn and Gln (+0.9840) and iTRAQ® isobaric tag Tyr (+304.2054) as variable modification. The false discovery rate (FDR) was estimated with decoy-fusion option included in the software. Proteins were then filtered with FDR < 1% (corresponding to a −10logP score above 25) for peptide-spectrum matches (PSMs) and a valid protein identification required minimum 2 unique peptides with a −10logP score above the peptide filtering threshold that can be mapped to only one protein group.

### 2.8.2. Protein iTRAQ® quantitation

The eight labeled samples (four replicates each of proteins from liquid stationary cultures and from halite brine inclusions) were mixed in equimolar ratios and injected in nanoLC-MS/MS in triplicate to reduce instrument variability as previously described. Quantitation was performed using PEAKS Q (quantitation program) iTRAQ 8-plex type with 10 ppm mass tolerance and only peptides with a score above FDR threshold 1% are used to quantify the identified proteins. Resulted quantitation was filtered accepting only proteins groups with fold change ≥2, at least two unique peptides and FDR adjusted ≤1% for both proteins identification and fold change significance using ANOVA significance method. Mass spectrometry proteomics data have been deposited to the ProteomeXchange Consortium via the PRIDE partner repository with the dataset identifier PXD037167 (project DOI: 10.6019/PXD037167).

### 2.8.3. Sample comparisons and functional annotation

Potential biases during analysis were avoided by filtering protein identifications produced using the PEAKS software to group proteins having multiple annotations in the NCBI database. To do so, we developed custom BASH and PYTHON scripts to merge all descriptions for one proteins identifier (favoring in order AAG-type, DAC-type, WP-type and then Q9H-type identifier). Venn diagrams were computed with VennDiagram and ggplot2 packages in R script. Functional proteins annotation was done with Kyoto Encyclopaedia of Genes and Genomes (KEGG, Kanehisa and Goto, 2000) using blastKOALA tool (Kanehisa et al., 2016; https://www.kegg.jp/blastkoala/), assigning KEGG Orthology identifier (also called K numbers) to identified proteins (see **Supplementary Table 6-2**). K number were then used for mapping identified proteins onto pre-existing metabolic pathways using KEGG Mapper search tools (https://www.kegg.jp/kegg/mapper/search.html; see **Supplementary Table 6-3**).

## 3. Results

### 3.1. Characterization of laboratory-grown halite containing *Halobacterium salinarum*

In order to study the physiology of *H. salinarum* cells after entombment within brine inclusions, we first needed to characterize the laboratory-grown halite, including resultant crystal size, cell localization, and surface-bound biomolecules. This enabled the establishment of selection criteria for downstream proteomics analyses and the development of protocols to exclude any "contaminant" proteins from the surface of the crystal.

### 3.1.1. Variability of *Halobacterium salinarum* cells from inoculated laboratory-grown halite

Laboratory-grown halite crystals produced from TNPA solutions containing *H. salinarum* cells using a slow evaporative process exhibited heterogeneous crystallization with respect to crystal size and coloration. As shown in **Figure 1**, even replicates produced under the same conditions produced different quantities of crystals of varying lengths and widths, but with relatively uniform thicknesses (±1 mm) of hopper and fishtail crystals. Color variations observed for halite in this study were likely the result of heterogeneities in the number of brine inclusions containing pigmented Halobacterium cells. The red color indicative of *H. salinarum* cells tended to concentrate in the center of hopper crystals where inclusions were most abundant (Roedder, 1984). Individual crystals of similar size and coloration were collected aseptically for further study, selecting only halite without visible defects signaling potential rupture of near-surface brine inclusions.

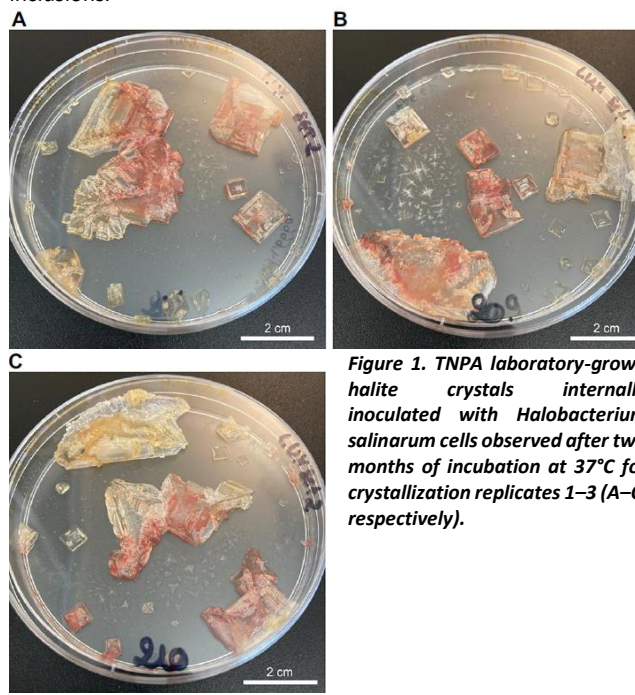

*Figure 1. TNPA laboratory-grown halite crystals internally inoculated with Halobacterium salinarum cells observed after two months of incubation at 37°C for crystallization replicates 1–3 (A–C, respectively).*

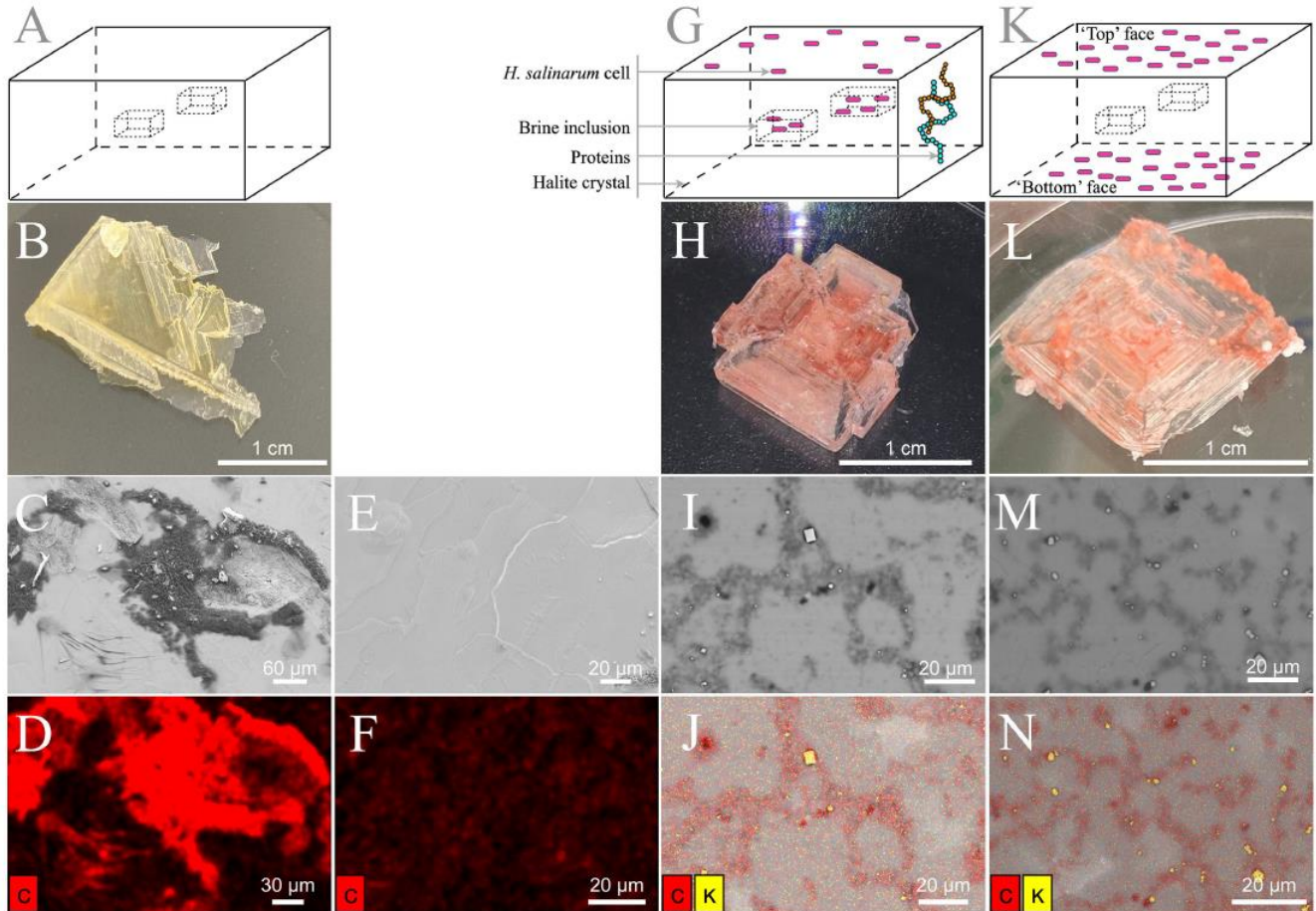

*Figure 2. Laboratory-grown halite surface observation. (A–F) TNPA crystals without cells. (G–J) Internally inoculated TNPA crystals with Halobacterium salinarum cells incubated 2 months at 37°C. (K–N) Externally inoculated TNPA crystals with Halobacterium salinarum cells after ~1 week incubated at 37°C. (A, G, K) Schematic representation of halite crystals without cells, internally inoculated and externally inoculated, respectively. (B, H, L) Visual observation of crystals without cells, internally inoculated and externally inoculated, respectively. (C, E) Secondary electron images. (I, M) Backscattered images. (D, F, J, N) Elemental composition analysis by EDX spectroscopy correlated, respectively, to the SEM images in panels C, E, I, M. Carbon and potassium appear in red and yellow, respectively.*

### 3.1.2. Contaminating surface-bound cells and proteins

To evaluate the potential for 'contaminating' microbial cells or residual biomolecules at the crystal surface, the elemental composition of halite with and without *H. salinarum* cells (**Figures 2A, B, G, H, K, L**) were determined by SEM-EDX. Elemental analyses of control halite derived from evaporation of the TNPA solution only exhibited the expected sodium chloride crystal surface, along with dense, irregular C agglomerates (**Figures 2C–F**) derived from the presence of organics (peptone, L-Arg) in the TNPA solution. The addition of *H. salinarum* cells to the TNPA solution prior to evaporation produced similar structures to control crystals, with no intact cells of *H. salinarum* visible on halite surfaces (**Figures 2I, J**). However, EDX analyses revealed differences in the distribution of carbon compared to control samples, as well as the presence of microscopic KCl crystals at the halite surface (**Figures 2I, B**) that were not found on control TNPA-derived crystals without *H. salinarum* cells. These differences were due to lysis of surface adhered *H. salinarum* cells releasing high concentration of cytosolic $K^+$ and $Cl^-$ accumulated as part of the "salt-in" osmoadaptation strategy (Engel and Catchpole, 2005). In order to assess both the potential for surface bound contaminants, including viable cells as well as proteins, externally inoculated halite were produced (**Figure 2K**). After these externally inoculated halite crystals were dissolved in growth medium and incubated at 37 °C, 180 rpm to assess viability. Such cultures reached a normal stationary phase culture ($OD_{600}$ > 1.0), demonstrating that at least a sub-population of halite surface-attached cells were viable. Additionally, SEM-EDX analysis of externally inoculated crystals revealed a similar elemental composition and distribution to that of internally inoculated halite (bearing cells both within brine inclusions as well as on the halite surface) after two months post-evaporation (**Figures 2M, N**). Taken together, these results confirm the presence of both viable cells and cellular debris on the surface of halite after two months of desiccation, validating the need for halite surface cleaning procedure to remove any residual proteins.

## 3.2. Development of total protein isolation methods for halite

For proteomics approaches, a new methodology was developed. This method was based on the protocol of Kirkland et al. (2006) using TRIzol reagent. Briefly, protocol was modified to optimize desalting steps (due to direct halite extraction with high NaCl content), protein pellet dissolution and protein digestion (see **Supplementary Information Section 4** for details on the parameters tested during protocol development and optimization). To ensure that only proteins contained within fluid inclusions are extracted using this protocol, it was also necessary to develop methods to remove contaminating proteins from halite crystal surfaces.

## 3.3. Development of proteolysis protocols for removal of halite surface-bound contaminants

Previous halite cleaning procedures from Gramain et al. (2011) and Sankaranarayanan et al. (2011) (see **Supplementary Table 1-1**) were developed to deactivate surface-contaminating microorganisms and remove surface-contaminating DNA. However, these methods were not designed for the removal of other types of biomolecule contaminants, such as proteins. A new, modified surface cleaning procedure was therefore needed to remove surface protein contamination to enable extraction of proteins only from within halite brine inclusions while avoiding crystal dissolution during the cleaning process. To meet these requirements, both cold plasma treatments and chemical protocols were tested using externally inoculated halite.

Cold plasma exposure is routinely used in astrobiology to sterilize spacecraft surfaces of microbial contaminants prior to launch. For this reason, a cold plasma approach was investigated here for the decontamination of halite surface. Two plasma sources – a DBD and an APPJ – were tested as an alternative to conventional liquid solvents. The expected benefit was to minimize halite dissolution-recrystallization events, hence ensuring the survival of the microorganisms within the halite. However, owing to their insulating properties and irregular topography, halite crystals attenuate plasma electric field and/or bend its field lines. As a result, and whatever the implemented plasma treatment (Ar, He, He mixed with $O_2$), the externally inoculated halite crystals still retained the pink coloration which is indicative of the protein and lipid pigments of *H. salinarum* cells (data not shown). As shown in **Table 1**, the most efficient plasma treatment drove to partial cell lysis (as determined by culture-based viability tests following growth by $OD_{600}$) and to a proteolysis of only 39.1% based on extraction and quantification of remaining proteins from halite surfaces using a modified BCA assay for low protein quantities; a value that remains insufficient for the application.

As an alternative, chemical treatments were tested, using protocols modified from the microorganism deactivation and DNA removal treatments of Sankaranarayanan et al. (2011) to extend to proteolysis. First, reduction of the exposure times in each chemical bath from previously published protocols were tested to avoid crystal dissolution, while maintaining the use of NaCl-saturated chemical solutions to avoid crystal dissolution. Subsequent, 5-min sequential NaOCl and NaOH treatments, either with or without additional HCl treatments, resulted in insufficient halite surface proteolysis (**Supplementary Figure 5-1a**). This was likely due to the rough structure of hopper crystals, which could allow proteins to remain attached during passive chemical baths. After chemical treatments partially degraded the contaminating biomolecules and reduced their attachment to the halite surface, an active-spray method was used for wash steps to increase the efficiency of biomolecule removal and prevent re-association with halite surfaces, compared to the use of passive wash baths (**Supplementary Figure 5-1b**). This active-spray method also allowed for both deactivation of halite surface-bound microorganisms (as indicated by a lack of culture growth in post-treatment viability tests) and better protein lysis compared to passive chemical bath washes. Using the protein extraction method described above, the efficacy of active-spray method for removal of surface-bound proteins was tested comparing NaOCl-NaOH and NaOCl-NaOH-HCl protocols. A shown in Figure 3A, addition of an acid wash results in a proteolysis efficiency of 93.0 ± 4.3%, compared to from of 83.7 ± 15.9% without the HCl treatment step (residual proteins 51.6 ± 31.6 µg and 120.5 ± 117.6 µg, respectively, n = 3). These results suggest that NaOCl-NaOH-HCl active-spray washes protocol (**Figure 3B**) present optimal and more consistent surface proteolysis, particularly for the small laboratory-grown halite crystals used in this study.

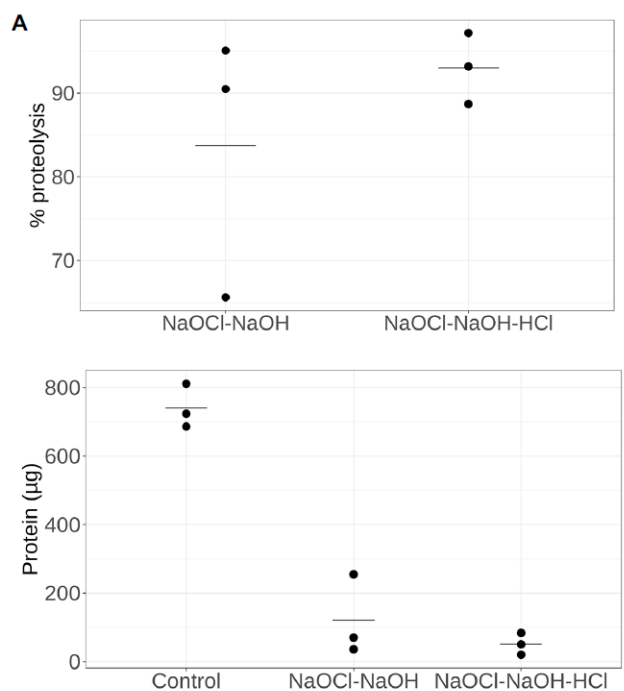

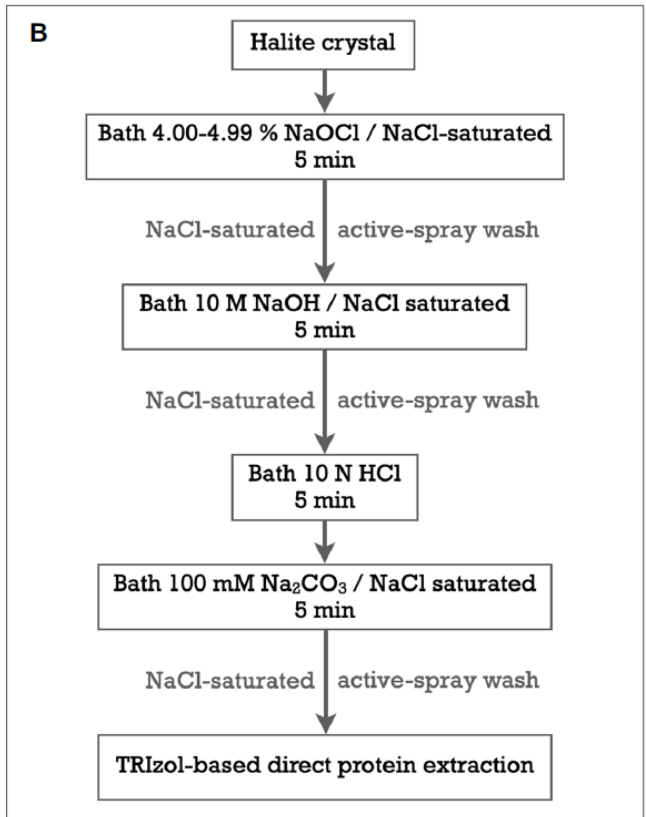

*Figure 3. Surface-bounded protein removal by chemical treatments with active-spray washes. (A) Percentage of proteins removed after NaOCl-NaOH treatment with or without an additional HCl treatment step, and subsequent residual protein quantity. (B) Overview of optimal NaOCl-NaOH-HCl chemical treatment with active-spray washes.*

To ensure that this active-spray chemical treatment did not affect *H. salinarum* cells within the halite brine inclusions, growth was monitored for crystals containing cells trapped in brine inclusions after surface chemical cleaning with active-spray NaOCl-NaOH-HCl treatment. Viable cultures were obtained from dissolution of the crystals in CM growth medium, thereby validating the active-spray chemical surface protein removal protocol for use in studying biomolecules within halite brine inclusions.

These new methods for removal of surface-bound proteins followed by extraction of total proteins from halite brine inclusions using a modified TRIzol-based method with desalting were then applied to answer the question of how *H. salinarum* acclimates to halite brine inclusions over the initial phase post-evaporation using proteomics.

## 3.4. Proteomic shifts in *Halobacterium salinarum* induced by acclimation to halite brine inclusions

Total proteins were extracted from the brine inclusions of halite after two months of complete dryness, with surface-bound proteins removed using the new protocol detailed above. These brine inclusions extracts were compared to total proteins extracted from *H. salinarum* cultures in stationary growth phase, representing conditions prior to halite formation. The proteins unique to each condition, as well as the common proteome between liquid cultures and from halite interiors were analyzed to determine how *H. salinarum* acclimates to halite brine inclusions.

Using the mass spectrometry data of four replicates from each condition, 1,249 total proteins were identified from stationary phase liquid cultures and 1,052 from halite brine inclusions, representing 1,342 unique proteins. Comparisons of the four replicate samples per condition revealed a core proteome composed of 839 proteins common to all sample replicates for liquid cultures and 715 proteins for halite brine inclusions extracts (**Figures 4A, B**). Of these, 655 proteins were expressed by cells both in liquid cultures and within halite brine inclusions (hereafter referred to as "shared proteins"); while 60 were specific to halite brine inclusions samples and another 184 were only identified from liquid cultures (**Figure 4C**; **Supplementary Table 6-1**). KEGG blastKOALA searches provided matches for roughly 75% of these shared proteins (488 of 655 shared proteins), as well as 62% of proteins unique to halite brine inclusions samples (37 of 60 proteins) and 52% of proteins unique to liquid cultures (95 of 184 proteins; see **Supplementary Table 6-2**), allowing for functional pathway reconstruction (see **Supplementary Table 6-3**).

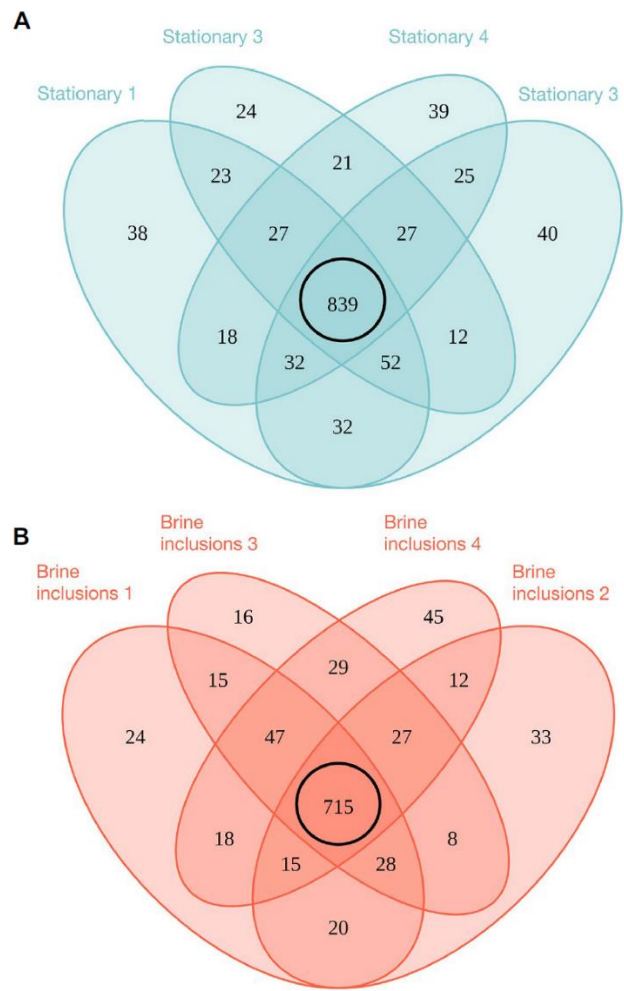

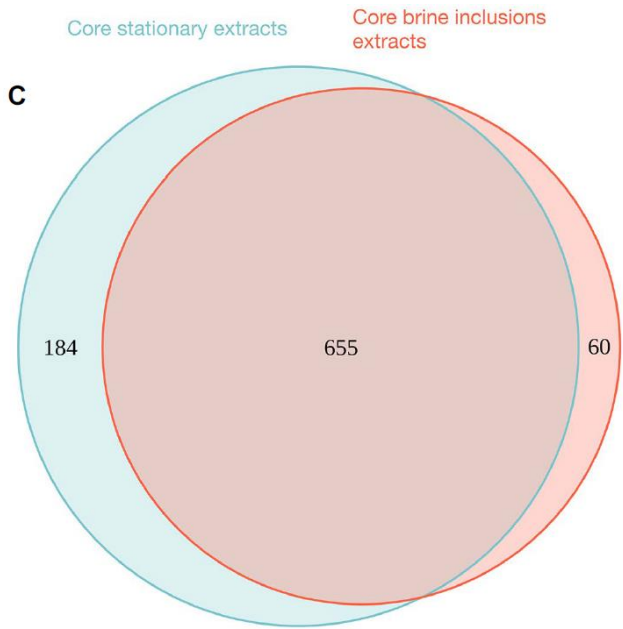 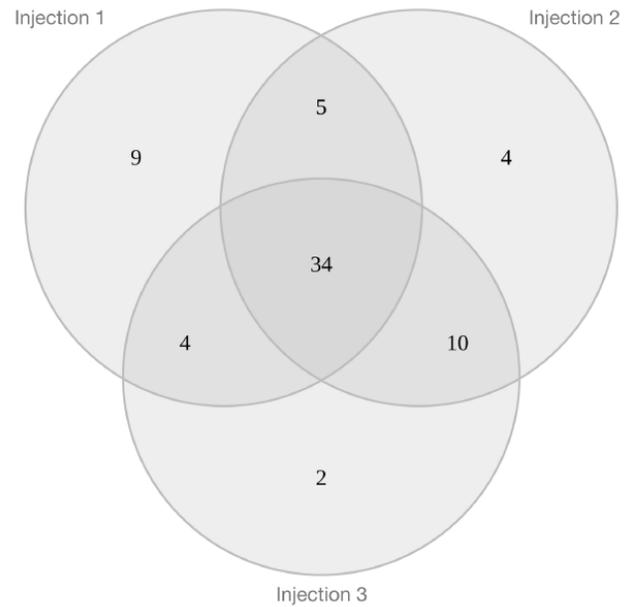

*Figure 4. Venn diagrams of proteins identified by mass spectrometry for brine inclusions and liquid stationary culture samples. (A) Comparison of biological replicates for liquid stationary culture extracts. (B) Comparison of biological replicates for brine inclusions extracts. Black circles in (A,B) represent "core" proteins shared by the four replicates in both cases. (C) Comparison of core proteins between brine inclusion extracts and liquid stationary culture cells.*

*Figure 5. Venn diagram of the 68 proteins quantified by iTRAQ® with triple injection which exhibit significative differential expression in brine extracts compared to liquid stationary culture cells.*

As a qualitative approach is insufficient to determine the regulation of proteins leading to their differential expression for *H. salinarum* cells within halite brine inclusions, a semi-quantitative mass spectrometry analysis was also performed. Results from all three injections were pooled to reduce instrument variability (**Figure 5**). Subsequently 68 proteins from halite brine inclusions were identified with statistically significant changes in expression levels compared to liquid cultures (fold change ≥2; see **Supplementary Table 6-4** and **Supplementary Figure 6-1**). KEGG blastKOALA searches provided functional pathway results for 80% of down-regulated proteins (35 of 44 proteins) and 71% of up-regulated proteins (17 of 24 proteins; see **Supplementary Table 6-2**).

A summary of the proteomics results is shown in **Figure 6** (and **Supplementary Table 6-5**), showing shared proteins between cells from the stationary growth phase cultures and brine extracts as well proteins differentially expressed (up- or down-regulated) in halite-derived samples. Investigations of the cell activity targeted proteins involved in central metabolism, energy production, cell division, replication, transcription and translation pathways. The cell interactions with the closed halite brine inclusions microenvironment were examined via cell envelope proteins (surface layer cell wall proteins and transporters) as well as proteins involved in the motility processes (chemotaxis, gas vesicle and archaellum).

### 3.4.1. Retention of proteins for central metabolism

Proteins involved in central metabolism, including key enzymes involved in glycolysis/gluconeogenesis and the TCA cycle, formed part of the shared proteome between halite brine inclusions extracts and stationary phase liquid cultures (**Figure 6**; **Supplementary Table 6-5**). Additionally, proteins involved in arginine fermentation and aerobic respiration were shared between halite brine inclusions and stationary growth phase liquid cultures. Quantitative analyses showed that the majority of proteins involved in pyruvate metabolism maintained consistent expression levels after two months of halite inclusion with the exception of up-regulation of proteins involved in acetyl-CoA production. Surprisingly, transporters and enzymes involved in ADI pathway were expressed by both cells in stationary phase liquid cultures and brine inclusions extracts, including the arginine deiminase, ornithine-carbamoyl transferase, carbamate kinase, arginosuccinase, arginosuccinate synthetase, arginine-ornithine antiporter and an IclR family transcriptional regulator. Only arginine deaminase was found significantly up-regulated for cells from halite brine inclusions. Proteins for key components of the electron transport chain, including NADH dehydrogenase/oxidoreductase subunits B, C, D, and I, NAD(P)/FAD-dependent oxidoreductase, succinate dehydrogenases and ATP synthase subunits A, B, C, D, E, H, and I were not found to be differentially regulated between cells from free-living (liquid cultures) or trapped within halite brine inclusions (with the exception of with F subunit lacked in one brine inclusions extract).

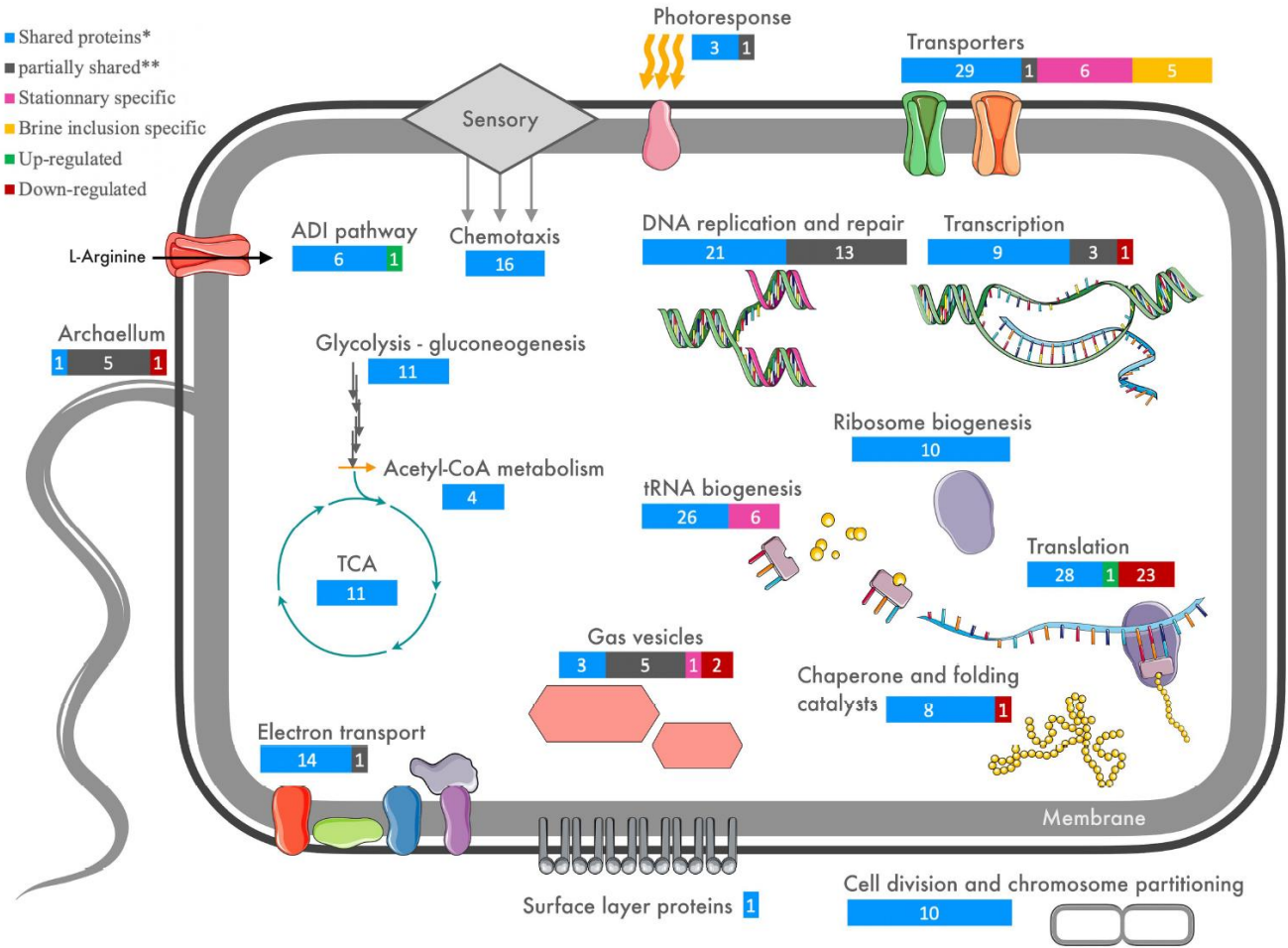

*Figure 6. Schematic overview of targets cellular functions involved in early acclimation of Halobacterium salinarum through halite brine inclusions (for extended matched proteins, see Supplementary Table 6-5). Shared (\*) indicates proteins found in all four brine inclusion extracts and all four liquid stationary culture extracts. Partially shared (\*\*) indicates proteins found in both brine inclusions and stationary phase liquid cultures, but not for all replicate samples for each condition.*

### 3.4.2. Similar expression of DNA repair, replication, and cell division proteins

Eight proteins involved in chromosome partitioning were common to halite brine inclusions extracts and stationary growth phase cultures, along with 11 other proteins involved in DNA replication including matches for DNA primase, polymerase, gyrase, topoisomerase and replication factors. Only four DNA replication proteins were found to be specific to stationary growth phase proteomes. DNA repair processes were represented both within halite brine inclusions and liquid cultures by the DNA photolyase, Rad50, the UvrD repair helicase, UvrA excinuclease, RadA recombinase, Rad25 DNA repair helicase, Fen1 repair endonuclease and the Hel208 helicase are found shared between liquids and brine extracts. Other DNA repair proteins were specific to either halite or liquid samples. Stationary growth phase cultures contained additional repair proteins including the RmeR site-specific deoxyribonuclease, endonuclease IV, RecJ single-strand exonuclease, MutL mismatch repair protein, DNA gyrase and N-glycosylase, whereas the UvrB, MutS and endonuclease III proteins were exclusive to halite brine inclusions proteomes.

### 3.4.3. Shared transcriptional but reduced translational proteomes

Investigations of DNA transcription identified a shared proteome between liquid cultures and brine inclusions including proteins for transcriptional initiation factors (TFIIE and TFIID) and transcriptional machinery (RNA polymerase subunits RpoA2, RpoB1, RpoB2, RpoD, and RpoE1). Proteins involved in transfer RNA biogenesis such as tRNA ligases for 18 different amino acids along with two ribonucleases completed the shared transcriptional proteome of liquid cultures cells and halite extracts. Only tRNA ligases specific for tryptophan and threonine, four unique transcriptional proteins (RpoF, RpoH, RpoN, and RusA termination protein) and one unique transcription initiation factor (TFIIB) were only found only in stationary growth phase cell extracts.

Translational activities were severely restricted for cell within halite brine inclusions, as evidenced by the down-regulation of 27

of the 42 ribosomal proteins shared by cells from both conditions tested. Only one ribosomal protein was up-regulated. The shared proteome between cells in halite brine inclusions and those in liquid culture included an additional 10 proteins involved in ribosome biogenesis, and 10 translation initiation factors for which none showed any significant up- or down-regulation. We attempted to corroborate the low ribosomal abundances by isolating RNA using the same TRIzol-based method (see **Supplementary Figure 2-1**). However, the RNA obtained from halite samples was of too low quality and quantity compared to that from liquid cultures (see **Supplementary Table 8-1**) for quantitative and transcriptional analyses. This was likely due to a combination of low RNA abundance in the halite fluid inclusions (based on ribosomal protein abundances) and some RNA degradation during extraction due to inefficient desalting using TRIzol alone (see **Supplementary Information Section 8** for further details). Chaperones proteins showed a high degree of conservation between conditions, with thioredoxin, chaperone DnaJ, DnaK along with the GrpE stimulator, Hsp20, thermosome subunit and prefoldin all shared by both liquid and halite extracts. Only prefoldin beta subunit was down-regulated for cells from halite brine inclusions.

### 3.4.4. S-layer maintained with minor cell envelope proteome changes in brine inclusions

S-layer proteins were identified in all liquid culture and halite samples without significant up- or down-regulation. Of the 20 membrane transporters shared between the proteomes of liquid cultures and cells inside halite brine inclusions, four were found to be up-regulated in halite brine inclusions extract, including the UgpB glycerol-3-phosphate-binding protein precursor. Moreover, six proteins were found to be unique to the halite brine inclusions proteome, with up-regulations of the phosphate, iron, peptide/nickel, and glycerol-3-phosphate transporters.

### 3.4.5. Modified sensory detection and motility

A high degree of variability was noted in the expression of these proteins between halite brine inclusions extracts and stationary growth phase cultures. Some proteins were common to both proteomes without any differences in protein quantities, including for the gas vesicle protein (Gvp) subunits F, C, N of cluster A, while GvpA was found only in two of the four liquid cultures and two of four halite samples. GvpD and I subunits of A cluster and F of B cluster were found punctually in liquid samples. However, the GvpO cluster A and B were significantly down-regulation in halite brine inclusions extracts.

The bacterio-opsin activator-like protein (Bat HTH-10 family transcription regulator; AAG 18778.1), along the 11 Htr signal transducers (including Htr-I and Htr-II that accompany the two sensory rhodopsins) were common to all conditions and replicates. In contrast, the bacterio-opsin activator (Bat; AAG19769.1) was only found in two of the four halite brine inclusions replicate samples. While this protein is hypothesized to be a DNA-binding transcriptional regulator in the photosensory network, its role remains unconfirmed. For the archaellum proteins, only archaellin B1 was down-regulated in halite samples whereas archaellin A1 and prearchaellin peptidase were identified punctually in liquid culture samples. Archaellin B2, A2 and flagellin related H protein were found in all liquid culture replicates but only in one halite replicate. An examination of chemotaxis proteins showed that the 18 identified proteins including CheR, CheA, CheW, CheC, CheB, and the methyl-accepting chemotaxis (MCP)-family proteins Htr transducers, were shared between halite brine inclusions extract and stationary growth phase liquid cultures.

# 4. Discussion

The initial objective of this study was to determine the molecular acclimation of haloarchaea to entrapment within halite brine inclusions using a proteomics approach. To this aim, we first needed to (1) better characterize the process of laboratory-based evaporation for halite entrapment of haloarchaea and its effects on both the resulting halite and the microbial cells, (2) develop efficient methods for the removal of halite surface-bound cells and biomolecules in order to isolate only proteins contained within the brine inclusions, (3) develop effective methods for protein extraction directly from halite brine inclusions in a shorter time period to avoid alterations in protein expression that can occur during gradual salt dissolution, and finally (4) combine these approaches to analyze total proteins from *H. salinarum* extracted from halite brine inclusions and compare these to control samples from stationary phase liquid cultures in order to propose a model of acclimation within halite. A slow evaporative process was chosen to generate the internally inoculated halite to model natural processes that occur over time scales sufficient for cellular acclimation to the environmental changes, rather than a rapid "shock" evaporative process.

## 4.1. Biotic/abiotic interactions influencing halite formation and *H. salinarum* viability

Similar to previous studies of survival of halophiles within halite brine inclusions, a NaCl-based buffered evaporation solution was used in this study (Norton and Grant, 1988; Fendrihan et al., 2012; Kixmüller and Greie, 2012). However, here we chose to add nutrient sources (1% peptone, 0.5% L-Arg HCl) to simulate organics present in the natural environment. This enabled us to study the role of metabolism in the survival and acclimation processes of *H. salinarum* cells to the early phases of entrapment within halite brine inclusions. The presence of microbial biomass and additional organics can affect halite precipitation processes and also explain the observed crystallization heterogeneity observed. For example, Norton and Grant (1988) previously demonstrated a positive correlation between the initial microbial cell density in liquid culture and the quantity of halite brine inclusions formed during evaporation.

The presence of nutrients seemed to diminish rather than increase the duration of survival. While Gramain et al. (2011) observed growth of *H. salinarum* NRC-1 after 27 months post-entombed in halite, there were some notable differences in the experimental parameters used. Gramain et al. (2011) used evaporation buffers with different compositions of nutrients to the evaporation buffer (either without nutrients or with addition of 0.01% Difco yeast extract and 0.0075% Merck casein hydrolysate diluted from the standard concentrations in modified Payne's medium of 1% and 0.75%, respectively (Payne et al., 1960)), incubated the resulting salt crystals in the dark at room temperature rather than with a 12 h:12 h diurnal cycle at 37°C as used here, and did not appear to employ halite surface sterilization or cleaning protocols prior to these survival tests. In contrast, our viability tests of *H. salinarum* cells from internally inoculated crystals incubated at 37°C over 80 days prior to subjected to surface organics cleaning treatments, dissolution and culturing showed no growth. The lower duration of survival observed for *H. salinarum* cells inside halite in our study may be the result of increased metabolic activity due to the higher concentration of nutrients in TNPA solution compared to the 100-time diluted modified Payne's medium used in Gramain et al. (2011). *H. salinarum* is capable of regulating changes in metabolic pathways in response to changes in carbon source availability (Schmid et al., 2009). This could result in an inhibitory effect due to the accumulation of metabolic waste products (Nyström, 2004) within the closed microenvironment of halite brine inclusions. The products of arginine fermentation include ornithine, $CO_2$ and $NH_4^+$. While the addition of peptone in this present study provided sufficient trace elements ($Mg^{2+}$, $K^+$, etc.) for nominal cell functions and S-layer stability at the moment of halite formation, these nutrients may be depleted over time within the closed environment of brine inclusions (Kixmüller and Greie, 2012), concurrent with the buildup of waste products. In contrast, Winters et al. (2015) showed that starvation of *H. salinarum* leads to smaller cell size, and hypothesized that this condition could contribute to extended survival within halite brine inclusions. This is somewhat counter-intuitive as natural evaporite environments contain the lysed remains of dead cells and other sources of organics.

On the other hand, in the absence of surface sterilization of halite used in the viability studies in Gramain et al. (2011), some of the surviving cells observed in the study may have been the result of surface-adhered cells rather than cells within the halite brine inclusions. These results also suggest a possible survival advantage for cells on halite surfaces rather than those within the brine inclusions over the early stages of evaporation and acclimation, a hypothesis supported by the findings of Gramain et al. (2011) showing no difference in growth for *H. salinarum* cells evaporated in salt buffer without nutrients and those containing the diluted Payne's medium nutrients. This epilithic lifestyle is likely supported in part by the organics, K, and Cl observed in this study by SEM–EDX analyses to accumulate on the halite surface, derived from the lysis of *H. salinarum* cells. It is important to note that the vacuum conditions for SEM observations may have resulted in rupture of unfixed *H. salinarum* cells, which may have otherwise remained intact under normal atmospheric conditions.

Altogether, these results of surface contamination confirm that the isolation of proteins exclusively from halite brine requires the removal of halite surface-bound cells and biomolecules.

## 4.2. Removal of halite surface-bound microorganisms and biomolecules

The small sizes of laboratory-grown halite proscribe the use of treatment processes that could result in salt dissolution. Cold atmospheric plasmas were therefore tested in an effort to avoid the use of liquid cleaners based on their effectiveness for the sterilization of spacecraft surfaces for planetary protection (Shimizu et al., 2014) and agricultural applications (Judée et al., 2018). Unfortunately, none of the plasma conditions tested in this study (gas mixtures, power delivery modalities) enabled a complete removal of proteins from the rough textured halite surfaces. However, the results remain encouraging insofar as cold plasma has demonstrated some effects on proteolysis that must now be amplified. Further experiments will be required to innovate an ad hoc plasma process delivering active species at higher densities while treating the whole surface of halite regardless their roughness surface, dimensions or dielectric permittivity.

The deactivation of microorganisms and removal of nucleic acids from halite surfaces were been therefore instead performed by adapting chemical wash methods first developed by Rosenzweig et al. (2000) and then optimized by Gramain et al. (2011) and Sankaranarayanan et al. (2011). By reducing the exposure times in sequential NaOCl, NaOH, and HCl treatments, and replacing passive chemical baths with an active spray process, we were able to achieve sufficient proteolysis of surface-bound proteins for downstream isolation of proteins exclusively from within halite brine inclusions, while avoiding halite dissolution during treatment. Compared to total proteins extracted from surface sterilized internally inoculated crystals (±1 mg order), residual surface bound proteins (51.6 µg) were marginal. However, observations of residual pigmentation after removal of halite surface-bound proteins suggests the presence of non-protein pigments such as carotenoids. Further refinements are therefore needed before future studies can isolate lipids exclusively from within halite brine inclusions.

## 4.3. Benefits and limitations of the TRIzol-based method for direct extraction of biomolecules from halite brine inclusions

TRIzol reagent has the distinct benefit of sequential separation of RNA, DNA, and proteins. TRIzol-based protein extraction methods have previously been used and validated to study proteins from liquid cultures of haloarchaea (Kirkland et al., 2006; Bidle et al., 2008). In this study, we developed and employed a modified procedure that allowed for direct protein extraction from salt crystals, with subsequent desalinations steps compatible with semi-quantitative mass spectrometry analyses from these high salt extracts. Most importantly, our approach avoids induced bias in proteome due to alterations in protein expression over the extended times needed for a slow crystal dissolution prior to protein extraction using other methods.

Although peripheral membrane proteins were identified using this approach, transmembrane peptides were absent in the mass spectrometry dataset. We hypothesize that this was due to retention of transmembrane protein domains with the lipid membrane fraction during TRIzol extraction in residual phenol phase after protein precipitation with isopropanol (see **Supplementary Figure 2-1**). Importantly, some transmembrane proteins could be identified by peptides outside the membrane-spanning domains, e.g., the ArcD arginine-ornithine antiporter protein, for which seven cytosolic- and extracellular-facing peptides were identified (see **Supplementary Figure 7-1**). However, not all transmembrane proteins were identified as evidenced by the lack of identified peptides for bacteriorhodopsin in any samples (stationary phase liquid cultures and brine inclusions). One plausible explanation of this phenomenon is the desalting effect of isopropanol (as suggested by Kirkland et al. (2006)) that can result in increased protein instability, leading to loss of conformation and subsequent peptide fragmentation for peptides outside the transmembrane regions. While these results indicate that further membrane disruption steps may be needed for complete extraction of all transmembrane proteins, similar to the protocol used by Podechard et al. (2018) to isolate membrane lipids using TRIzol, the potential bias introduced by our methodology is limited, as evidenced by the identified peptides on either side of the lipid membrane.

Moreover, while TRIzol-based methods have been used for RNA extraction from liquid cultures of *H. salinarum* (De Lomana et al., 2020), it was shown here to be produce RNA from halite fluid inclusions of insufficient quality for RNA-Seq. Indeed, it would appear that further desalting steps would be needed and optimized for extraction of high-quality RNA (see **Supplementary Information Section 8** for further details).

## 4.4. Postulate of cellular origin of total brine extracted proteome

The proteins extracted using the protocol presented here represent the total protein complement of the halite brine inclusions. These proteins may be components of viable or non-viable cells, or even proteins released by cells into the extracellular environment of the brine inclusions. Many questions remain about the potential for salts to preserve proteins over time as molecular biosignatures, particularly within the protected evaporite brine inclusions microenvironment. Considering that viability tests performed on the internally inoculated halite crystals demonstrate cell viability after two months of *H. salinarum* entrapment, we postulate here a cellular origin for the extracted proteins.

## 4.5. Acclimation of *Halobacterium salinarum* to halite brine inclusions

We applied the slow laboratory evaporation method to *H. salinarum* cultures in TNPA buffer followed by removal of halite surface-bound cells and organics (particularly proteins) and selective extraction of proteins from within halite brine inclusions described in the preceding sections. This allowed us to study the early (two months) acclimation of the haloarchaeal cells to halite brine inclusions using a proteomics approach. Stationary growth phase cultures were used as a control to approximate the cell physiology prior to a slow evaporative process. Our analyses were focused on two main themes: (1) cell activity, and (2) interactions between cells and local microenvironment within the brine inclusion. Halite brine inclusions are enclosed microenvironments. While questions remain about possible alterations of brine inclusion composition over geological time scales, during the initial phase after crystal formation the composition is based on the initial hypersaline environment. Conditions of near-saturating salt concentrations lead to low dissolved oxygen available to cells. Cells are therefore hypothesized to be in a physiological state similar to stationary growth phase, with low cell division and altered metabolic activity. Here we examined proteome alterations in response to acclimation to this unique microenvironment.

### 4.5.1. Cell activity within halite brine inclusions

The metabolism of Halobacterium cells trapped within brine inclusions is hypothesized to shift from aerobic metabolism by respiratory chain to anaerobic fermentation through arginine deiminase pathway (ADI) or ATP generation via photoheteroorganotrophy. Halite incubations were performed in this study with a 12 h:12 h light:dark photoperiod, allowing for both phototrophy via bacteriorhodopsin and arginine fermentation under oxygen-limited conditions (Hartmann et al., 1980), as are presumed to exist inside closed brine inclusions. These two ATP-generating pathways are antagonistic for liquid cultures of *H. salinarum*, but can theoretically alternate over the diurnal cycle used here to simulate surface conditions two months after halite precipitation. However, exposure to light also depends on the location of the brine inclusion within the halite crystal, with inclusions near the crystal surface receiving higher total irradiance than inclusions near the crystal center. While the extraction protocol used enabled identification of phototrophy-related proteins such as bacteriorhodopsin activators, it did not allow for extraction of many membrane proteins including bacteriorhodopsin itself. Thus, it is not possible to determine if phototrophy-related proteins were differentially expressed in liquid and halite samples. Under anaerobic conditions, Hartmann et al. (1980) showed that while retinal biosynthesis was inhibited by the absence of oxygen, even low levels of bacteriorhodopsin could produce appreciable levels of ATP. A basal expression of proteins for different ATP generating pathways could enable the survival of *H. salinarum* in the early stages of halite entombment. Additionally, the potential exists for biomolecule recycling within brine inclusions, similar to that demonstrated by prokaryotes in deep subsurface (Thomas et al., 2019) or other resource-limited environments to limit energy consumption linked to biosynthesis. The reduced viability of *H. salinarum* in brine inclusions in the presence of organics including arginine compared to starvation conditions seems to favor fermentative metabolism leading to the accumulation of metabolic waste products. Confirmation of the precise metabolic pathways used by *H. salinarum* within halite brine inclusions will require further analyses, overcoming the challenges of accessing the closed halite system.

Acclimation to the halite brine inclusions microenvironment could presumably involve reduced or silenced cell division, DNA replication and repair pathways. However, the majority of proteins detected implicated in genome maintenance were not differentially expressed between halite brine inclusions and stationary growth phase cultures. This suggests that either such proteins are constitutively expressed, or that DNA replication and cell division occur in a similar fashion for both stationary growth phase cells and those within halite brine inclusions. While early acclimation to the halite brine inclusions microenvironment resulted in nuanced differences in the proteome of *H. salinarum* cells regarding cell division, replication, DNA repair and transcriptional processes, stark differences were observed for proteins involved in translation activities. Among the shared proteome showing no differential expression between conditions were proteins involved in DNA replication, reparation and transcriptional pathways. This suggests similar levels of genome maintenance between late growth stage cultures and brine entombment. However, translational proteins were not detected in brine extracts, potentially indicating decreased de novo protein synthesis compared to stationary cells. These seemingly contradictory results can be explained by transcriptional activity directed not to mRNA synthesis but to regulatory RNA. However, this hypothesis is not fully satisfactory as tRNA synthesis and ribosomal biogenesis protein quantities remain similar for both conditions.

Taken together, these data suggest a model for cellular activity during early acclimation to halite brine inclusion highly similar to that of stationary growth phase cells in liquid culture. Active metabolism appears to continue, with data suggesting anaerobic fermentation. However, it is important to note that the extraction method used had limited success in isolating membrane proteins such as bacteriorhodopsin and sensory rhodopsins. Therefore, their absence in this data set cannot be taken as confirmation of absence in the cell, and as such diurnal cycling between anaerobic metabolisms (arginine fermentation in the dark and photoheteroorganotrophy in the light) cannot be ruled out. It is important to note as well that proteomic analyses considered all brine inclusions of a given crystal in a bulk analysis that did not distinguish between brine inclusions near the crystal surface receiving higher irradiance from those deeper within the crystal. Cells maintain at least the capacity for central metabolism, DNA repair, replication, and even cell division comparable to stationary growth phase cells. Acclimation to the closed brine inclusions microenvironment appears to severely limit translational activity. However, expressed proteins are preserved indicating low levels of protein turnover. This raises the question of how cells are able to interface with the surrounding microenvironment within the inclusions.

### 4.5.2. Interactions between cells and the brine inclusions microenvironment

The potential interactions between *H. salinarum* cells and the local microenvironment within halite brine inclusions focusing on cell surface and mobility processes involved in responses to external stimuli. Previous investigations on the survival of Halobacterium cells trapped within salt crystals have proposed potential for cell envelope modifications with surface proteinaceous layer (S-layer) shed (Fendrihan et al., 2012). However, this result seems to be the due to a lack of trace elements in the TN buffer (4.28 M NaCl, Tris–HCl pH 7.4) used in the 2012 study, as argued by Kixmüller and Greie (2012). The peptone added to the TNPA crystallization buffer used here provided sufficient trace elements (including Mg, Ca, K) for S-layer maintenance and cell envelope function. S-layer shedding within halite brine inclusions was previously thought to improve exchanges between cells and the local microenvironment. However, in this study S-layer proteins were identified in all liquid culture and halite samples without significant up- or down-regulation. This either indicates the S-layer is maintained by *H. salinarum* cells, or that the S-layer proteins were preserved within the brine inclusion. The fact that S-layer proteins quantities remained comparable to stationary phase cultures with intact S-layers suggests that the S-layer proteins remain associated with *H. salinarum* cells within brine inclusions.

The increased expression of membrane transporters suggests that transport capacity is not only maintained but enhanced in cells during acclimation to halite brine inclusions, likely to maximize extraction of essential molecules and trace elements from the local microenvironment. This may represent a form of biomolecule recycling similar to that demonstrated by procaryotes in deep subsurface (Thomas et al., 2019) or other resource-limited environments to limit energy consumption linked to biosynthesis. The up-regulation of the glycerol-3-phosphate-binding protein precursor may suggest a role for membrane lipid modification within brine inclusions.

Cell motility could logically be assumed to be of lesser importance for *H. salinarum* cells inside the restricted microenvironment of a brine inclusion. The small volume of such inclusions limits the heterogeneity of available oxygen and nutrients, as well as constraining the ability of cells to move away from potentially damaging elements. However, the detection of the Htr signal transducers of the sensory rhodopsin I and II as part of the common proteome of all conditions tested indicates that despite being constrained to brine inclusions, the phototaxis remains functional both prior to and after evaporation of *H. salinarum* cultures.

In contrast, down-regulations of the GvpO cluster A and B were observed for halite brine inclusions extracts. A similar down-regulation of gvpO mRNA was observed in response to UV irradiation (Baliga et al., 2004), and GvpO protein expression levels after exposure to ionizing radiation (Webb et al., 2013). Lack of the GvpO gas vesicle expression regulator is thought to inhibit gas vesicle formation, and growth by anoxic arginine fermentation is also known to reduce the number of gas vesicles in *H. salinarum* NRC-1 (Hechler and Pfeifer, 2009). Therefore, the lack of gas vesicle production could be linked to anaerobic metabolism. Gas vesicle down-regulation suggests a lack of needed capacity for vertical movement in a water column in response to oxygen levels, consistent with the restricted confines of a microscopic halite brine inclusions. This correlates with previous published transcriptomics and proteomics studies on *H. salinarum* cell showing the impact of salinity and dissolved oxygen concentrations on motility (Schmid et al., 2007; Leuko et al., 2009).

If the sensory pathways leading to cell motility remain intact, the proteins required for motility itself are absent. Our data show a general down-regulation of archaellin proteins involved in motility for cells in brine inclusions compared to stationary phase liquid cultures. So, while chemo- and photo-taxis proteins (including Che-family proteins and Htr signal transducers) are maintained in brine inclusions, the accompanying archaellum motility is not. Reduced motility may be induced in halite brine inclusions to preserve energy. The expression of multiple transducer proteins is often correlated with changing environmental conditions, and in this case may be the remanent proteins from the acclimation of *H. salinarum* to halite brine inclusions.

# 5. Conclusion

This is the first study that demonstrate the possibility to isolate proteins directly and efficiently from halite brine inclusions while simultaneously excluding surface-bound contaminating proteins and avoiding changes in cell protein expression during salt dissolution. Also, this work provides the first insights into the molecular mechanisms involved in the early acclamation of *H. salinarum* within brine inclusions. Based on our findings, the cells of *H. salinarum* appear to be in a low activity "maintenance" or "semi-dormant" state, similar to other bacteria and archaea in low-energy environments such as the deep biosphere (reviewed in Lever et al. (2015)). Whether or not *H. salinarum* cells inside brine inclusions are true "persister" cells (Megaw and Gilmore, 2017) will require future study. Further investigation is also needed to explore the hypothesis of biomolecules recycling similar to that observed in deep sediments (Thomas et al., 2019). Despite the reasonable assumption brine inclusions represent stress conditions for Halobacterium cells, no increased expression of stress-response proteins such as chaperones was observed. The acclimation to the halite environment may therefore be less a stress response than a reduction in cell activity.

Shifts in the proteome of *H. salinarum* NRC-1 have been studied using liquid cultures under conditions similar to those thought to exist inside halite brine inclusions: reduced oxygen content (Schmid et al., 2007), increased salinity (Leuko et al., 2009), and transitions from aerobic to anaerobic growth (Tebbe et al., 2009). While the down-regulation of certain ribosomal proteins and archaellum precursors was also observed following salinity stress (Leuko et al., 2009) not all protein expression shifts conformed to this pattern. The same was observed for the transition from aerobic to anaerobic growth, with increased expression of the arginine deaminase as observed by Tebbe et al. (2009) but other protein expression differences due to the use of culturing conditions that were not identical to those in previous studies. Thus, while proteomics analyses from liquid cultures can provide hints to aid in the interpretation of the data presented here, they cannot fully explain the observed patterns of protein expression within halite brine inclusions.

The unique environment of halite brine inclusions limits exchange of nutrients between cells and their surrounding environment. It also allows for the buildup of metabolic waste products which appear to limit the duration of cell viability within the brine inclusions. Further functional investigations are clearly needed to confirm the hypotheses generated by proteomics data in this study, particularly in regard to cell activity. While this study does not resolve the question of how long microorganisms are able to retain viability within halite brine inclusions, it offers clear insights into the process of early microbial acclamation to evaporation of hypersaline environments in halophilic archaea. The methodologies presented here will also enable future study of the biomolecules of microorganisms in halite inclusions in natural settings, particularly small-sized halite crystals.

**Data availability statement**
The datasets presented in this study can be found in online repositories. The names of the repository/repositories and accession number(s) can be found in the article/Supplementary material.

**Author contributions**
Study was designed by AK in collaboration with CF, with the participation of AH and SZ in experimental design and supervision. Experiments were conducted by CF with the help of MM for crystals forming. AM and RP for mass spectrometry and improvements of extraction methods. BA-B for nanoLC-MS/MS data acquisition. FG for MEB-EDX analyses. AT for RNA extraction and TD for plasma treatments. Proteomics data treatments were conducted by AT and CF. Manuscript was written by CF. All authors contributed to the article and approved the submitted version.

**Funding**
This work was supported by the X-life program of CNRS-MITI, the ATM program of the Muséum National d'Histoire Naturelle, the French National Research Agency ANR-PRCI "ExocubeHALO" project (ANR-21-CE49-0017-01_ACT), and the Sorbonne Université (graduate stipend CF).

**Acknowledgments**
UHPLC MS/MS data were acquired at the Plateau Technique de Spectrométrie de Masse Bio-organique, UMR 7245 Molécules de Communication et d'Adaptation des Microorganismes, Muséum national d'Histoire naturelle, Paris, France. The nanoLC-MS/MS data were acquired at ProGenoMix MS-platform, IBiSA labeled, at CEA/SPI/Li2D. Thanks to Imène Esteve of the FIB and SEM facility of IMPMC which was supported by Région Ile de France Grant SESAME 2006 NOI-07-593/R, Institut National des Sciences de l'Univers (INSU)–CNRS, Institut de Physique–CNRS, Sorbonne Université, and the French National Research Agency (ANR) grant ANR-07-BLAN-0124-01. Parts of figures used images from Servier Medical Art. Servier Medical Art by Servier is licensed under a Creative Commons Attribution 3.0 Unported License (https://creativecommons.org/licenses/by/3.0/).

**Conflict of interest**
The authors declare that the research was conducted in the absence of any commercial or financial relationships that could be construed as a potential conflict of interest.
Publisher's note